\documentclass[twocolumn,prb,aps,showpacs]{revtex4}
\usepackage{graphics}
\begin{document}
\title{\bf Fermi gas with attractive potential and arbitrary spin in one-dimensional trap}
\vspace{1.5em}
\author{P. Schlottmann$^1$ and A.A. Zvyagin$^{2,3}$}
\affiliation{$^1$ Department of Physics, Florida State University, Tallahassee, Florida 32306 \\
$^2$ B.I. Verkin Institute for Low Temperature Physics and Engineering, \\
Ukrainian National Academy of Sciences, 47 Lenin Avenue, Kharkov, 61103, Ukraine \\
$^3$ Max-Planck-Institut f\"ur Physik komplexer Systeme, D-01187, Dresden, Germany}
\date{\today}
\begin{abstract}
A gas of ultracold $^6$Li atoms (effective spin 1/2) confined to an elongated trap with one-dimensional properties is a candidate to display three different phases: (i) fermions bound in Cooper-pair-like states, (ii) unbound spin-polarized particles, and (iii) a mixed phase which is believed to have some resemblance to the FFLO pairing. It is of great interest to extend these studies to fermionic atoms with higher spin, e.g., for neutral $^{40}$K, $^{43}$Ca, $^{87}$Sr or $^{173}$Yb atoms. Within the grand-canonical ensemble we investigated the $\mu$ vs. $H$ phase diagram for $S=3/2$ ($\mu$ is the chemical potential and $H$ the external magnetic field) for the ground state using the exact Bethe {\it ansatz} solution of the one-dimensional Fermi gas interacting with an attractive $\delta$-function potential. There are four fundamental states: The particles can be either unpaired or clustered in bound states of two, three and four fermions. The rich phase diagram consists of these four states and various mixed phases in which combinations of the fundamental states coexist. Bound states of four fermions are not favorable in high magnetic fields, but always present if the field is low. Working within the grand-canonical ensemble has the following advantages: (1) A universal phase diagram is obtained by scaling with respect to the interaction strength and (2) possible scenarios for phase separation are explored within the local density approximation. The phase diagram for the superposition of a Zeeman and a quadrupolar splitting is also discussed. 
\end{abstract}

\maketitle

\section{Introduction}

Spin-imbalanced ultracold $^6$Li atoms confined to different geometries are spin-1/2 fermion systems displaying the interplay of Cooper pairing and spin-polarization and have been the subject of several recent studies.\cite{Zwierlein,Shin,Partridge} Confinement to nearly one-dimensional tubes can be achieved if the ultracold cloud of atoms is subjected to a two-dimensional optical lattice, which defines a two-dimensional array of tubes.\cite{Liao} The tubes can be regarded as isolated if the confinement by the laser beams is strong enough to suppress tunneling between tubes. The scattering between atoms under transverse harmonic confinement is subject to a confinement-induced resonance.\cite{Olshanii} Fine-tuning this Feshbach-type resonance, the interaction between the fermions can be made attractive and its strength can be varied.\cite{Bergeman} The interaction is local and can be approximated by a $\delta$-function potential in space.  The confinement along the tube is roughly harmonic and weak; it can be locally incorporated into the chemical potential. Consequently, these systems of fermions are only locally homogeneous and within the local density approximation display phase separation with the variation of the chemical potential along the tube.\cite{Orso,Hu} 

One-dimensional spin-1/2 gases have been extensively studied theoretically. M. Gaudin\cite{Gaudin} and C.N. Yang\cite{CNYang} extended Bethe's {\it ansatz} for the Heisenberg chain\cite{Bethe} and Lieb and Liniger's results for the locally interacting gas of bosons\cite{Lieb} to obtain the exact solution for a gas of spin-1/2 fermions interacting via a $\delta$-function potential. It was shown by Gaudin\cite{Gaudin} and later by Takahashi\cite{Takahashi} and Lai\cite{Lai} that for an attractive interaction in the ground state there are two classes solutions of the discrete Bethe {\it ansatz} equations, namely, real charge rapidities and paired complex conjugated rapidities. The former represent spin polarized particles and the latter correspond to bound states of the Cooper type.  There are then three possible homogeneous phases at very low $T$, the (1) fully spin-polarized state (only real charge rapidities), (2) a phase without polarization, where all particles are bound in Cooper-pairs, (only complex conjugated rapidities), and (3) a mixed phase in which unpaired spin-polarized particles coexist with Cooper pairs. The Cooper pairs are gapped (i.e. it requires a critical field to break-up the bound states) and display no long-range order. Similar results were obtained for the Hubbard model with attractive $U$.\cite{BWoynarovich,LeeS}  The mixed phase has been interpreted\cite{Yang} as the one-dimensional analogue of the Fulde-Ferrell-Larkin-Ovchinnikov (FFLO) state.\cite{FFLO} For an isolated tube there is no long-range order of the pairs and hence no order parameter; however, the Cooper-pair correlation function acquires a phase in the mixed state, that is believed to be reminiscent of the space modulation of the order parameter in higher dimensions. A coupling between tubes increases the effective dimension of the system so that long-range order can arise and it is believed that this could lead to the realization of the FFLO phase in an ultracold gas of atoms.\cite{Liao,Yang} FFLO related phases have been observed in the strongly anisotropic heavy electron compound\cite{Radovan,Bianchi} CeCoIn$_5$ (the interpretation is still controversial\cite{Kenzelmann}) and in the quasi-two-dimensional organic compounds $\lambda$-(BETS)$_2$FeCl$_4$ and (TMTSF)$_2$ClO$_4$.\cite{Uji,Yonezawa} 

Tubes with ultracold gases of atoms provide the unique possibility to study fermion systems with a spin larger than 1/2, e.g., $^{40}$K (spin 9/2), $^{43}$Ca (spin 7/2), $^{87}$Sr (spin 9/2) or $^{173}$Yb (spin 5/2) atoms. With an attractive interaction atoms with spin $S$ can form bound states of up to $(2S+1)$ particles, extending this way the concept of Cooper pairs to larger clusters.\cite{He} Consequently, the phase diagram will have more possible pure and mixed phases. In this paper we investigate the phases that can arise in the ground state using the Bethe {\it ansatz} solution of the one-dimensional fermion gas with $\delta$-function potential. Sutherland\cite{Sutherland} generalized M. Gaudin's\cite{Gaudin} and C.N. Yang's\cite{CNYang} Bethe {\it ansatz} solution (for spin 1/2) to an arbitrary number of colors $N = 2S+1$ [SU($N$)-symmetry]. The classification of states and the thermodynamics of the gas have been derived long ago by one of the present authors\cite{Schl93} and the ground state equations and elementary excitations have been previously studied in Refs. [\onlinecite{GuYang,Schl94}], for both attractive and repulsive potential and arbitrary number of colors. Introducing different chemical potentials for each of the colors, these results are valid for an arbitrary level splitting of the $N$-fold multiplet.\cite{Schl93,Schl94,Schl97} The results for the $\delta$-function potential model, as well as other integrable models, have been extensively reviewed by one of the present authors.\cite{Schl97} In this paper we use the solution to study the phase diagram for a gas of fermionic atoms constrained to a tube. To address a concrete example here we consider $S=3/2$; the numerical effort and the complexity of the phase diagram rapidly increase with $S$. For $S=3/2$ there are fours basic states, namely, bound states of four, three and two particles, and unbound particles, and the corresponding mixed phases, which can have up to four coexisting basic states. As a consequence of Pauli's exclusion principle, the bound states must involve particles all with different spin-components; otherwise the wave function has a node and the $\delta$-function potential cannot be active.

Recently, in the context of ultracold atoms, the Australian National University group\cite{Guan1,Guan2,Guan3,PengHe} reconsidered the SU($N$) $\delta$-function potential model with attractive interaction. In Ref. [\onlinecite{Guan1}], using the thermodynamic Bethe {\it ansatz}, the dressed energy potential are re-derived for "spin-1 fermions", and used to discuss the phase diagram for a superposition of a Zeeman and a quadrupolar splitting. In Ref. [\onlinecite{Guan2}], these results are extended to spin-3/2 fermions for fixed density, strong attractive coupling (Tonks-Girardeau gas limit) and Zeeman as well as quadrupolar splittings. This case makes contact with the present paper. The ground state energy, critical fields and the low-$T$ specific heat for the SU($N$) case are obtained in Ref. [\onlinecite{Guan3}]. The thermodynamics for the $SU(3)$ case has been studied in Ref. [\onlinecite{PengHe}]. Most results of these four papers\cite{Guan1,Guan2,Guan3,PengHe} are already derived in Refs. [\onlinecite{GuYang,Schl93,Schl94,Schl97}]. 

Note that the two-body interaction for spin larger than 1/2 does not necessarily have to have SU($N$) symmetry. Integrable one-dimensional continuum models for the low-density limit displaying pairing have been constructed for bosonic and fermionic systems. In Ref. [\onlinecite{Cao}] a model for spin-1 bosons with exchange interaction is proposed and solved exactly via nested Bethe {\it ans\"atze} for the ground state and thermodynamics. An extension of this model to $SO(5)$ symmetry for spin-3/2 fermions has been proposed and solved in Ref. [\onlinecite{Jiang1}]; the authors obtain the thermodynamic equations and discuss the spectrum of elementary excitations. Further extensions to models with hidden $Sp(2s+1)$ and $SO(2s+1)$ symmetries for high spin-$s$ fermions and bosons, respectively, can be found in Ref. [\onlinecite{Jiang2}]. The influence of a pure quadratic Zeeman effect (quadrupolar splitting) on the Mott-insulator phases of hard-core one-dimensional spin-3/2 fermions has been studied via DMRG, leading to a rich phase diagram.\cite{Rodriguez}   

There are several other theoretical studies of ultracold spin-1/2 atoms in one-dimension. The direct imaging of the density profiles of the spatially modulated superfluid phases in atomic fermion systems were obtained by solving the Bogoliubov-de Gennes equation.\cite{Mizushima} The pairing states were investigated on a lattice by means of the density matrix renormalization group method in Ref. [\onlinecite{Feiguin}]; this study leads to a fourth possible phase (in addition to the paired, unpaired polarized and the mixed phases) consisting of a metallic shell with free spin-down (i.e. reversed spins) fermions moving in a fully filled background of spin-up fermions. The crossover from three-dimensional (FFLO phase) to one-dimensional (mixed phase) behavior is addressed in [\onlinecite{Parish}], where the phase diagram for a weakly interacting array of tubes is calculated. A quantum Monte Carlo study of one-dimensional trapped fermions with attractive contact interactions was presented in [\onlinecite{Casula}]. Finally, using the Bethe {\it ansatz} the low temperature thermodynamics was calculated in Refs. [\onlinecite{Guan}] and [\onlinecite{Bolech}].   

The rest of the paper is organized as follows. In Sect. II we present the model and the discrete Bethe {\it ansatz} equations for perodic and open boundary conditions.  The elongated traps correspond to open boundaries rather than to the usually employed periodic boundary conditions. This difference gives rise to surface or boundary terms, which, however, are not relevant in the dressed energy potentials in the thermodynamic limit. In Sect. III we present the numerical solution of the Bethe {\it ansatz} equations, the phase diagram for a Zeeman splitting and the local density profile along the trap. In Sect. IV we investigate the case where in addition to the Zeeman effect there is a quadrupolar splitting. Although it is not clear if nonlinear Zeeman splittings are of relevance to ultacold atoms in one dimension, it is an instructive situation to study which has been considered in Refs. [\onlinecite{Cao,Guan1,Rodriguez}].  Conclusions are presented in Sect. V. 

\section{Model and Bethe {\it ansatz}}

The Hamiltonian for a gas of nonrelativistic particles with $(2S+1)$ colors (spin $S$) interacting via an attractive $\delta$-function potential is 
\begin{equation} 
{\cal H} = - \sum_{i=1}^{N_p} \frac{\partial^2}{\partial x_i^2} - 2 |c| \sum_{i < j} \delta(x_i - x_j) \ , \label{H}
\end{equation}
where $x_i$ are the coordinates, $N_p$ is the total number of particles and $c$ is the interaction strength. By fine-tuning the confinement-induced resonance\cite{Olshanii} the interaction can become attractive and its strength can be varied. Here $\hbar^2/2m$, where $m$ is the mass of the particles, has been equated to $1$, or alternatively it has been scaled into ${\cal H}$ and $c$.  

\subsection{Bethe equations for periodic boundary conditions} 

The states of the coordinate Bethe {\it ansatz} are plane waves constructed from the two-particle scattering matrix. This scattering matrix satisfies the so-called Yang-Baxter triangular relation, which is a necessary condition for integrability. As a consequence of the triangular relation many-particle scattering processes can be factorized into two-particle processes and the order in which the individual scattering processes take place can be interchanged (the order becomes arbitrary). 

The generalization of the Gaudin-Yang\cite{Gaudin,CNYang} solution to more than two colors\cite{Sutherland} consists of an iterative application of the Bethe-Yang hypothesis (generalized Bethe {\it ansatz}), such that one color is eliminated at each step, leading to $N=2S+1$ nested Bethe {\it ans\"atze}. Each Bethe {\it ansatz} gives rise to a new set of rapidities, $\{k_j\}$, $j=1,\cdots,N_p$ for the charges (coordinate Bethe {\it ansatz}) and $\{\Lambda_{\alpha}^{(l)}\}$, $l=1,\cdots,N-1$, with $\alpha=1,\cdots,M^{(l)}$ for the internal degrees of freedom (spin).  Here $M^{(l)}$ is the number of rapidities in the $l^{th}$ set and $\alpha$ is the running index within each set.  All rapidities within a given set have to be different to ensure linearly independent solutions.  Consider fermions of spin $S$ with Zeeman splitting and let us denote by $N_{S-m}$ the number of particles with spin component $m$.  We have then $N_{S-m_1} \ge N_{S-m_2}$ if $m_1 > m_2$ and define
\begin{equation}
M^{(i)} = \sum_{m=-S+i}^S N_{S+m} \ , \ M^{(0)} = N_p \ , \ M^{(2S+1)} = 0 \ , \label{M}
\end{equation}
such that $N_p \ge M^{(1)} \ge \cdots \ge M^{(2S)} \ge 0$. As a consequence of the $SU(N)$ invariance of the model the nested Bethe {\it ans\"atze} for {\bf periodic} boundary conditions yield the following sets of coupled equations\cite{Sutherland,Schl93,Schl97}
\begin{eqnarray}
&&\exp\bigl(i k_j L\bigr) = \prod_{\beta=1}^{M^{(1)}} e\bigl(k_j - \Lambda_{\beta}^{(1)}\bigr) \ , \ j = 1 , ... , N_p  \\
&&\prod_{\beta=1}^{M^{(l-1)}} e\bigl(\Lambda_{\alpha}^{(l)} - \Lambda_{\beta}^{(l-1)}\bigr) \prod_{\beta=1}^{M^{(l+1)}} e\bigl(\Lambda_{\alpha}^{(l)} - \Lambda_{\beta}^{(l+1)}\bigr) \nonumber \\
&&= - \prod_{\beta=1}^{M^{(l)}} e\bigl[\bigl(\Lambda_{\alpha}^{(l)} - \Lambda_{\beta}^{(l)}\bigl)/2\bigr] \ , \ \alpha = 1 , ... , M^{(l)} \nonumber \\
&&l=1,\cdots,2S \label{periodic} 
\end{eqnarray}
where
\begin{equation}
e(x) = \frac{x - i {\textstyle \frac{1}{2}}|c|}{x + i {\textstyle \frac{1}{2}}|c|} \ , \label{e}
\end{equation}
$\Lambda_j^{(0)} \equiv k_j$ and $L$ is the length of the box. The energy and the momentum of the state are given by
\begin{equation}
E = \sum_{j=1}^{N_p} k_j^2 \ , \ P = \sum_{j=1}^{N_p} k_j \ . \label{EP}
\end{equation}

\subsection{Bethe equations for open boundary conditions}

Eqs. (\ref{periodic}) are derived for the standard periodic boundary conditions. Tubes, however, are not periodic and better represented by {\bf open} or reflecting boundary conditions. A particle reaching the boundary is then reflected undergoing $k_j \to -k_j$ but without changing its energy. The corresponding reflection matrix satisfies {\it reflection equations} with the two-particle scattering matrix, extending the Yang-Baxter equations. All matrices can be diagonalized simultaneously.\cite{Sklyanin,Fendley} The total length of a period is now $2L$, where $L$ is the length of the trap. The open boundary Bethe equations for the present model were derived previously by Oelkers {\it et al}.\cite{Oelkers} It is convenient to write the Bethe equations in a form similar to Eqs. (\ref{periodic}) by letting the indices $j$ and $\alpha$ run from $-N_p$ to $N_p$ and $-M^{(l)}$ to $M^{(l)}$, respectively.\cite{FrahmZvyagin,Zvyagin} The Bethe Ansatz equations for {\bf open} boundary conditions are then
\begin{eqnarray}
&&\exp\bigl(i 2k_j L\bigr) e\bigl(k_j\bigr) = \prod_{\beta=-M^{(1)}}^{M^{(1)}} e\bigl(k_j - \Lambda_{\beta}^{(1)}\bigr) \ , \nonumber \\
&&j = -N_p , ... , N_p \label{open2} \\
&&e\bigl[\bigl(\Lambda_{\alpha}^{(l)}/2\bigr)\bigr] \prod_{\beta=-M^{(l-1)}}^{M^{(l-1)}} e\bigl(\Lambda_{\alpha}^{(l)} - \Lambda_{\beta}^{(l-1)}\bigr) \nonumber \\
&&\times \prod_{\beta=-M^{(l+1)}}^{M^{(l+1)}} e\bigl(\Lambda_{\alpha}^{(l)} - \Lambda_{\beta}^{(l+1)}\bigr) = - \left[e\bigl(\Lambda_{\alpha}^{(l)}\bigr)\right]^2 \nonumber \\
&&\prod_{\beta=-M^{(l)}}^{M^{(l)}} e\bigl[\bigl(\Lambda_{\alpha}^{(l)} - \Lambda_{\beta}^{(l)}\bigr)/2\bigr] \ , \ \alpha = -M^{(l)} , ... , M^{(l)} \ , \nonumber \\
&&l = 1,\cdots,2S \ . \label{open3} 
\end{eqnarray}
Hence, there are twice as many rapidities and the box is also twice as large, leaving the density of rapidities unchanged. The main difference between open and periodic boundary conditions are then the independent factors in Eqs. (\ref{open2}) and (\ref{open3}), which contribute with $1/L$ terms to the rapidity densities. This is very similar to the effect of magnetic impurities in a chain. Also, for periodic boundary conditions the Bethe states are plane waves, while for open boundary conditions they are standing waves.
The energy and momentum are still given by Eq. (\ref{EP}).
 
\subsection{Classification of states and energy potentials}

For an attractive interaction and large $L$, the solutions of the discrete Bethe equations can be classified according to (i) real charge rapidities, belonging to the set $\{k_j\}$, associated with unpaired propagating spin-polarized particles, (ii) complex spin and charge rapidities, which correspond to bound states of particles with different spin components, and (iii) strings of complex spin rapidities, which represent bound spin states.\cite{Schl93,Schl97} States in class (iii) are not represented in the ground state; these states correspond to excited states and are not considered here.  This classification of states is completely analogous to that of the Anderson impurity of arbitrary spin in the $U \to \infty$ limit\cite{Schl84,Schl89} and the one-dimensional degenerate supersymmetric $t-J$ model.\cite{Schl87} 

Since only particles with different spin components are scattered, i.e. experience an effective attractive interaction, we may build bound states of up to $(2S+1)$ particles. A bound state of $n$ ($n \le N =2S+1$) is characterized by one real $\xi^{(n-1)}$ rapidity and in general complex $\Lambda^{(l)}$ rapidities, $l < n-1$, given by
\begin{eqnarray}
&&\Lambda_p^{(l)} = \xi^{(n-1)} + ip|c|/2 \ \ , \ \ l \le n-1 \le 2S \ \ , \nonumber \\
&&p=-(n-l-1),-(n-l-3),\cdots,(n-l-1) \ . \label{strings}
\end{eqnarray}
These spin and charge strings form classes (i) and (ii), which are present in the ground state.\cite{Schl93} The real rapidities $\xi^{(n-1)}$ have all to be different and satisfy the Fermi-Dirac statistics, i.e. the states are either occupied or empty. (For $S=1/2$ the bound states are frequently called Cooper pairs, although this analogy is not rigorous.) In the ground state the rapidities are densely distributed in the interval $[-B_l,B_l]$ and we denote with $\varepsilon^{(l)}(\xi)$, $l=0,1,\cdots,2S$ the dressed energy potentials (entering the Fermi-Dirac distribution). For $S=3/2$ the four energy potentials satisfy the following coupled linear integral equations (for larger spin see Refs. [\onlinecite{Schl93,Schl94,Schl97}]):
\begin{eqnarray}
&&\varepsilon^{(0)}(\xi) = \Bigl[\xi^2 - \mu_0\Bigr] - \int_{-B_1}^{B_1} d\xi' \varepsilon^{(1)}(\xi') a_1(\xi-\xi') \nonumber \\
&&- \int_{-B_2}^{B_2} d\xi' \varepsilon^{(2)}(\xi') a_2(\xi-\xi') \\ \label{var0}
&&- \int_{-B_3}^{B_3} d\xi' \varepsilon^{(3)}(\xi') a_3(\xi-\xi') , \nonumber \\ 
&&\varepsilon^{(1)}(\xi) = 2 \Bigl[ \xi^2 - \frac{1}{4}c^2 - \mu_1 \Bigr] \nonumber \\
&&- \int_{-B_0}^{B_0} d\xi' \varepsilon^{(0)}(\xi') a_1(\xi-\xi') \nonumber \\
&&- \int_{-B_1}^{B_1} d\xi' \varepsilon^{(1)}(\xi') a_2(\xi-\xi') \\ \label{var1}
&&-\int_{-B_2}^{B_2} d\xi' \varepsilon^{(2)}(\xi') \Bigl[a_1(\xi-\xi') + a_3(\xi-\xi') \Bigr] \nonumber \\
&&-\int_{-B_3}^{B_3} d\xi' \varepsilon^{(3)}(\xi') \Bigl[a_2(\xi-\xi') + a_4(\xi-\xi') \Bigr] , \nonumber \\ 
&&\varepsilon^{(2)}(\xi) = 3 \Bigl[ \xi^2 - \frac{2}{3}c^2 - \mu_2 \Bigr] \nonumber \\
&&- \int_{-B_0}^{B_0} d\xi' \varepsilon^{(0)}(\xi') a_2(\xi-\xi') \nonumber \\
&&- \int_{-B_1}^{B_1} d\xi' \varepsilon^{(1)}(\xi') \Bigl[ a_1(\xi-\xi') + a_3(\xi-\xi') \Bigr] \\ \label{var2}
&&-\int_{-B_2}^{B_2} d\xi' \varepsilon^{(2)}(\xi') \Bigl[a_2(\xi-\xi') + a_4(\xi-\xi') \Bigr] \nonumber \\
&&-\int_{-B_3}^{B_3} d\xi' \varepsilon^{(3)}(\xi') \Bigl[a_1(\xi-\xi') + a_3(\xi-\xi') + a_5(\xi-\xi') \Bigr] , \nonumber \\
&&\varepsilon^{(3)}(\xi) = 4 \Bigl[ \xi^2 - \frac{5}{4}c^2 - \mu_3 \Bigr] \nonumber \\
&&- \int_{-B_0}^{B_0} d\xi' \varepsilon^{(0)}(\xi') a_3(\xi-\xi') \nonumber \\
&&- \int_{-B_1}^{B_1} d\xi' \varepsilon^{(1)}(\xi') \Bigl[ a_2(\xi-\xi') + a_4(\xi-\xi') \Bigr] \\ \label{var3}
&&-\int_{-B_2}^{B_2} d\xi' \varepsilon^{(2)}(\xi') \Bigl[a_1(\xi-\xi') + a_3(\xi-\xi') + a_5(\xi-\xi') \Bigr] \nonumber \\
&&-\int_{-B_3}^{B_3} d\xi' \varepsilon^{(3)}(\xi') \Bigl[a_2(\xi-\xi') + a_4(\xi-\xi') + a_6(\xi-\xi') \Bigr] , \nonumber 
\end{eqnarray}
where
\begin{equation}
a_n(x) = \frac{1}{\pi} \frac{n|c|/2}{x^2 + n^2 c^2 /4} \label{an}
\end{equation}
and $\mu_l$ is the chemical potential for the bound states involving $(l+1)$ particles. The $\mu_l$ determine the integration limits $B_l$ through the condition that $\varepsilon^{(l)}(\pm B_l) = 0$, since occupied states correspond to $\varepsilon^{(l)}(\xi)<0$ and for empty states the quantity is positive. For a pure Zeeman splitting the chemical potentials are $\mu_0 = \mu +3H/2$, $\mu_1 = \mu +H$, $\mu_2 = \mu + H/2$ and $\mu_3 = \mu$, where $H$ is the Zeeman energy and $\mu$ the chemical potential for the total number of particles.\cite{Schl94} 

Note that if all the $\varepsilon^{(l)}$ are rescaled to $\varepsilon^{(l)}/c^2$, $\mu$ to $\mu/c^2$, $H$ to $H/c^2$, all $B_l$ to $B_l/|c|$ and $\xi$ to $\xi/|c|$, the equations are {\bf universal}, i.e., independent of the magnitude of $|c|$. Hence, within the framework of the grand canonical ensemble, without loss of generality, it is sufficient to present the results for $|c|=1$. The problem has then only two parameters, namely, $H$ and $\mu$. 

An ultra-cold atom system is inherently inhomogeneous since the diameter of the tube gradually changes with position from the center of the trap to its boundaries.  As a consequence of the changing diameter of the tube, the quantization in the plane transversal to the tube gradually changes the zero of energy.  This change can be represented by a harmonic potential, so that the actual local chemical potential $\mu$ is a function of $x$ given by
\begin{equation}
\mu(x) + {\textstyle \frac{1}{2}}m \omega_{ho}^2 x^2 = const \ . \label{mu}
\end{equation}
Within the local density approximation, it is $\mu(x)$ that enters the Bethe equations, (\ref{var0})-(\ref{var3}). The solution is then exact for the one-dimensional system, but approximative for the trap. This approximation\cite{Liao,Orso,Hu} is expected to be good since the variation of $\mu$ with $x$ is slow. The approximation neglects the quantization of the harmonic confinement, which is treated classically and locally incorporated into the chemical potential.

Fluctuations in the particle density arise due to the $x$-dependence of $\mu$ and due to possible weak Josephson tunneling between tubes. It is then necessary to solve the Bethe equations in the grand canonical ensemble rather than for the canonical ensemble, i.e. at constant number of particles. The results in the canonical ensemble may then no longer be universal, i.e. the scaling with $|c|$ discussed above is no longer valid, because of the constraint of fixed number of particles, which is the case treated in Ref. [\onlinecite{Guan2}].  

\section{Phase diagram for Zeeman splitting}

In this section we discuss the phase diagram for a pure Zeeman splitting of the levels. The set of equations (\ref{var0})-(\ref{var3}) is solved numerically by iteration. In this case, the energy potential $\varepsilon^{(0)}$ corresponds to unpaired particles with spin-component $S_z = 3/2$; the energy $\varepsilon^{(1)}$ to bound pairs with spin-components $S_z = 3/2$ and $S_z = 1/2$; the potential $\varepsilon^{(2)}$ to bound states of three particles of spin components $S_z = 3/2$, $S_z = 1/2$ and $S_z = -1/2$, respectively; and finally $\varepsilon^{(3)}$ to bound states of four particles all with different spin-components. We denote these states with Roman numbers, I, II, III, and IV, respectively. These states can coexist in mixed phases, for example we denote with I+IV the coexistence of unpaired and bound states of four particles and with I+II+III a phase where all states except four-particle bound states are present.  

\begin{figure}[!ht]
\begin{center}
\resizebox{0.49\textwidth}{!}
{\includegraphics{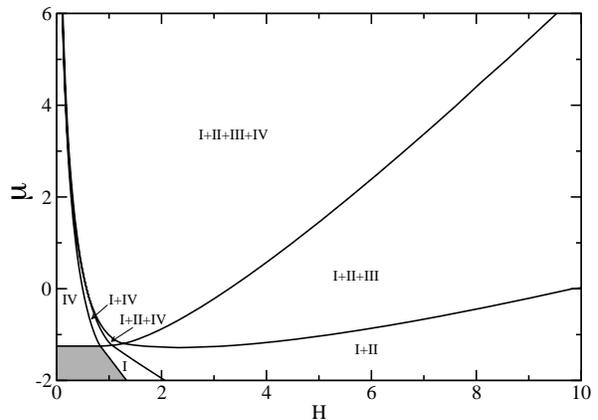}}
\end{center}
\caption{Ground state phase diagram $\mu$ vs. $H$ for a homogeneous fermion 
gas of spin $S=3/2$ with $|c|=1$. The shaded area corresponds to the empty system (no 
particles). The Roman numbers denote the number of particles involved in a 
bound states. Regions with more than one Roman number have coexisting or mixed
phases with coexisting order. Note that in the vertical axis $\mu$ is a function 
of $x$ as given by Eq. (\ref{mu}).}
\label{phasediagrFig1}
\end{figure}

\begin{figure}[!ht]
\begin{center}
\resizebox{0.45\textwidth}{!}
{\includegraphics{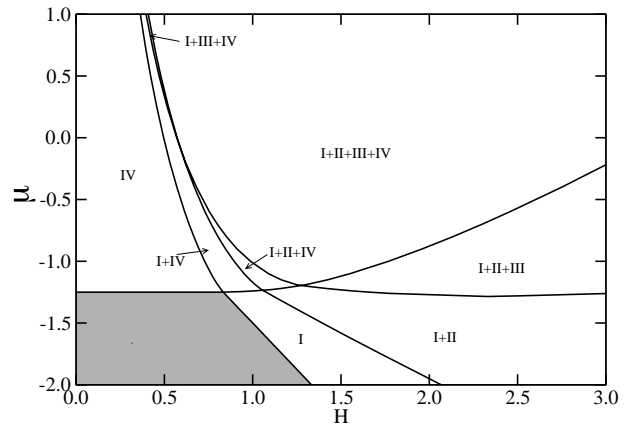}}
\end{center}
\caption{Expanded view of the ground state phase diagram $\mu$ vs. $H$ of Fig. 1
for a homogeneous fermion gas of spin $S=3/2$ with $|c|=1$ to show the multiple phases for
the low-density region. The shaded area corresponds to the empty system (no 
particles).} \label{phasediagrFig2}
\end{figure}

The phase diagram of $\mu(x)$ vs. $H$ for $|c|=1$ is shown in Fig. 1. Other values of $|c|$ can be reduced to this phase diagram by adequately scaling $\mu$ and $H$. Note that all phase boundaries are given by the zero of some energy potential. The phase boundaries are then crossover lines, which are accompanied by a square-root singularity of one of the densities of states (one-dimensional van Hove singularity).\cite{vanHove} For small magnetic fields all particles are bound in four-particle bound states (generalized Cooper pairs). The shaded area is the region where all bands are empty (system without particles). With increasing field other phases become realized. At very large magnetic fields and/or for low values of $\mu$ (small number of particles) the phase IV is not favorable. For large $\mu$ and intermediate magnetic fields all for bands are populated and hence spin-polarized unbound particles coexist with all possible bound states. This phase diagram can be compared to the one obtained in Ref. [\onlinecite{Guan2}] (Figs. 3 and 4) for fixed number of particles (canonical ensemble). For a pure Zeeman splitting and as a function of field these authors obtain crossovers from phase IV to I+IV to I. A constant number of particles corresponds to a curve $\mu(H)$ in Fig. 1. The strong interaction limit considered in Ref. [\onlinecite{Guan2}] corresponds to a low particle density. The sequence of phases we then expect from Fig. 1 is also IV to I+IV to I, in agreement with Ref. [\onlinecite{Guan2}].  

Fig. 2 shows the expanded view of the region for small $\mu$ and $H$, which displays multiple crossovers. For larger $\mu$ there is a small region where the phase I+III+IV is stable. As mentioned above, the harmonic confinement of the trap can be treated quasi-classically and can be absorbed into the chemical potential via Eq. (\ref{mu}).\cite{mu}  The chemical potential then decreases as we move from the center of the trap towards the boundaries. Hence, we move downward along a vertical line on the phase diagram. This can give rise to phase separation along the length of the trap. For instance, for $H=2$ for a sufficiently high density of atoms, at the center of the trap the phase with all bound states coexisting (I+II+III+IV) would be favored, then moving towards the end-points (in either direction) of the trap first the four-particle bound states disappear (phase I+II+III), then the bound states of three particles (phase III) are depopulated and polarized unbound particles coexist with bound pairs (phase I+II), and finally a fully polarized gas phase (I) is possible. For $H=1$, on the other hand, we again could have the I+II+III+IV phase at the center of the trap and by moving to the boundaries we would observe the I+II+IV mixed phase, then the I+IV phase and finally unbound fully polarized atoms (I). 

\vskip 0.3in
\begin{figure}[!ht]
\begin{center}
\resizebox{0.45\textwidth}{!}
{\includegraphics{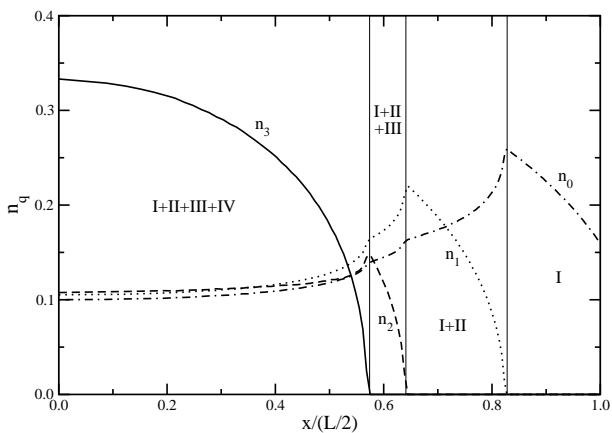}}
\end{center}
\caption{Density profile within the local density approximation for $H=1.5$, $\mu(0)=-0.7$ and $\mu(L/2)=-2.0$. The position along the trap is given by Eq. (\ref{mu1}). The three crossovers between phases are shown by the thin vertical lines.  The densities $n_q$ of bound states of $q+1$ particles (or polarized unbound particles if $q=0$) are given by the solid ($n_3$), dashed ($n_2$), dotted ($n_1$) and dash-dotted ($n_0$) curves.} 
\label{phasediagrFig2}
\end{figure}

The local density profile as a function of $x$ for the different phases for $H=1.5$ is displayed in Fig. 3, where Eq. (\ref{mu}) was used to parametrize the chemical potential in a trap of length $L$. Given $\mu(0)$ and $\mu(L/2)$, i.e. the chemical potential at the center and boundary of the trap, the position along the trap is given by (from Eq. (\ref{mu})) 
\begin{equation}
x/(L/2) = \sqrt{[\mu(x)-\mu(0)]/[\mu(L/2)-\mu(0)]} \ . \label{mu1}
\end{equation}

The density function of the rapidities is obtained from the dressed energies $\varepsilon^{(q)}(\xi)$ by differentiation with respect to $\mu$, i.e.,\cite{Schl94,Schl97}
\begin{equation}
\rho_h^{(q)}(\xi) + \rho^{(q)}(\xi) = - \frac{1}{2 \pi} \frac{\partial \varepsilon^{(q)}(\xi)}{\partial \mu} , \label{rho}
\end{equation}
where $\rho^{(q)}(\xi)$ is the particle density and $\rho_h^{(q)}(\xi)$ the corresponding hole density for bound states involving $q+1$ particles. The integral equations satisfied by the density functions is similar to the one for the dressed energy potentials. After solving these equations numerically, the number of bound states (or polarized unbound particles if $q=0$) per unit length is obtained from
\begin{equation}
n_q = \int_{-B_q}^{B_q} d\xi \rho^{(q)}(\xi) \ , \label{nq}   
\end{equation}
Note that the densities vanish with a square-root singularity that is characteristic of one-dimensional van Hove singularities as can be seen in Fig. 3.

\section{Phase diagram for Zeeman and quadrupolar splitting}

\begin{figure}[!ht]
\begin{center}
\resizebox{0.45\textwidth}{!}
{\includegraphics{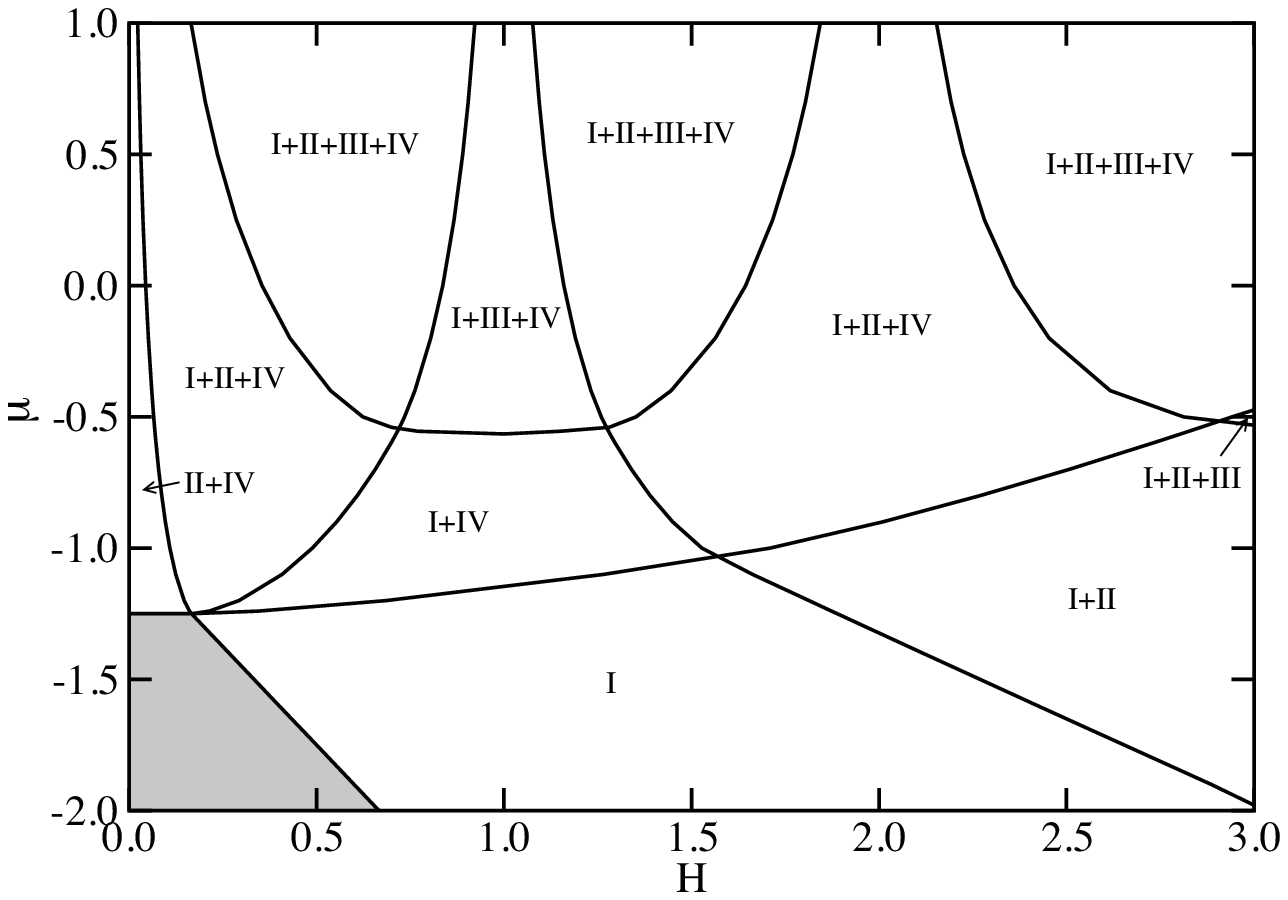}}
\end{center}
\caption{Ground state phase diagram $\mu$ vs. $H$ for a homogeneous fermion 
gas of spin $S=3/2$ with $|c|=1$ and a quadrupolar splitting $D[3S_z^2-S(S+1)]$ 
for $D=1/3$. The shaded area corresponds to the empty system (no particles). Level 
crossings of the Zeeman and quadrupolar terms occur at $H=3D$ and $H=6D$.} 
\label{phasediagrFig3}
\begin{center}
\resizebox{0.45\textwidth}{!}
{\includegraphics{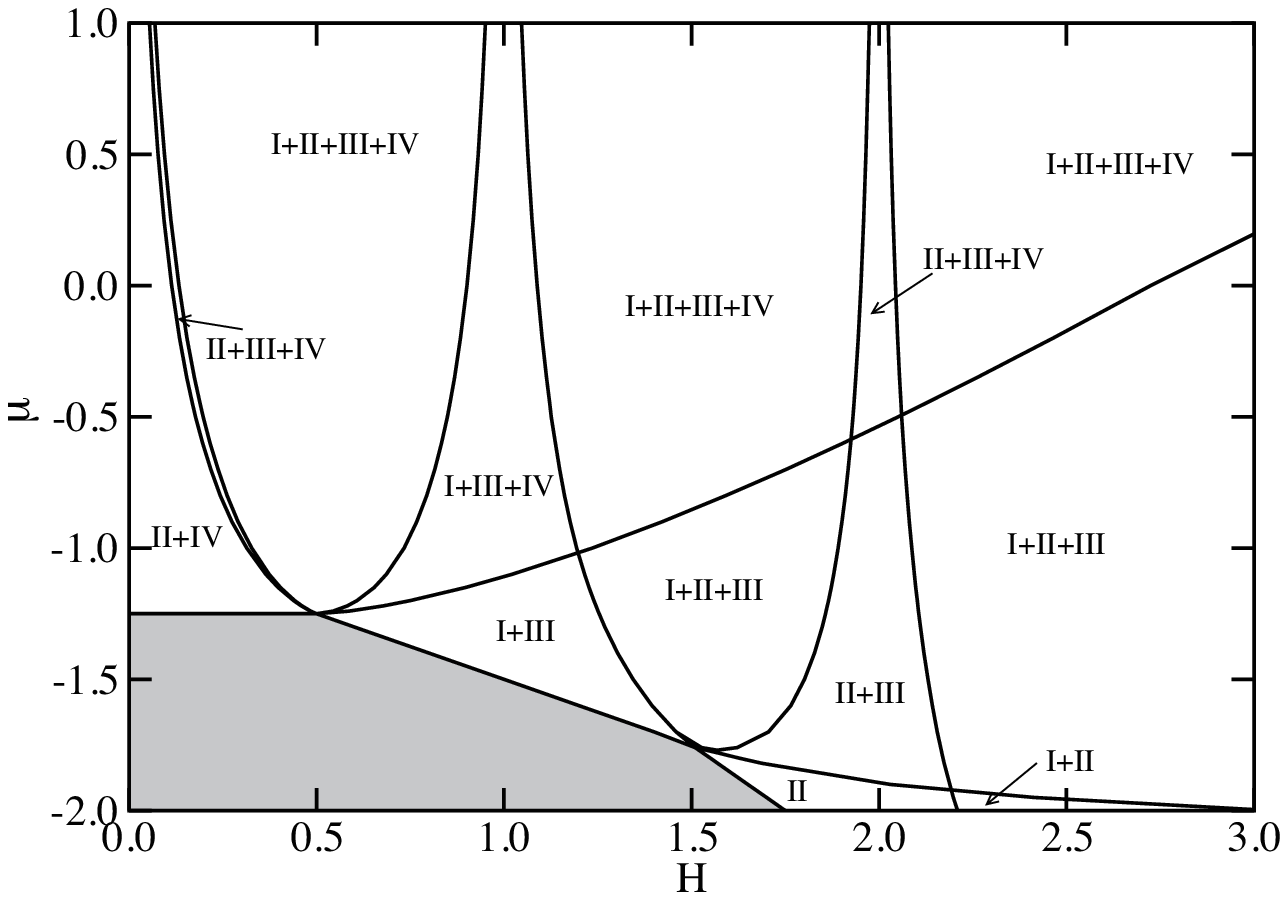}}
\end{center}
\caption{Ground state phase diagram $\mu$ vs. $H$ for a homogeneous fermion 
gas of spin $S=3/2$ with $|c|=1$ and a quadrupolar splitting $D[3S_z^2-S(S+1)]$ 
for $D=-1/3$. The shaded area corresponds to the empty system (no particles). 
Level crossings of the Zeeman and quadrupolar terms occur at $H=3|D|$ and $H=6|D|$.} 
\label{phasediagrFig4} 
\end{figure} 

For $S > 1/2$ level splittings other than the Zeeman effect are possible.  Here we consider a quadrupolar splitting superimposed with the Zeeman effect. Nonlinear Zeeman splittings have been considered previously\cite{Rodriguez,Guan1,Guan2} in similar contexts. In particular, in Refs. [\onlinecite{Guan1,Guan2}] Bethe {\it ansatz} equations for the same model but a fixed number of particles (canonical ensemble) were employed, while in the present paper we use the grand canonical ensemble (variable number of particles).  For $S=3/2$ the Zeeman and quadrupolar splittings are not the most general case, since also octupolar splittings are possible. It is not clear if such a situation (quadrupolar and octupolar splittings) can be realized experimentally for ultracold atoms in 1D, but the problem is theoretically sufficiently interesting to be addressed. The Hamiltonian for the level splitting is given by
\begin{equation}
{\cal H}_{spl} = -H S_z +D[3S_z^2 -S(S+1)] \ , \label{Ham}
\end{equation}
where $D$ can be either positive or negative. To be specific we consider again the case $S=3/2$. Note that the Hamiltonian (\ref{H}) commutes with ${\cal H}_{spl}$, so that the Bethe states also diagonalize ${\cal H}+{\cal H}_{spl}$. Keeping $D$ fixed and as a function of $H$, ${\cal H}_{spl}$ displays two level crossings; hence, we need to consider  three regions, namely, Region (i) ($H \le 3|D|$), Region (ii) ($3|D| \le H \le 6|D|$) and Region (iii) ($6|D| \le H$). For $D > 0$ the bound states I, II, III and IV are then composed by particles with the following $S_z$ components: In Region (i), (1/2), (1/2,-1/2), (1/2,-1/2,3/2), and (1/2,-1/2,3/2,-3/2), respectively; in Region (ii), (1/2), (1/2,3/2), (1/2,3/2,-1/2), and (1/2,3/2,-1/2,-3/2), respectively; and in Region (iii), (3/2), (3/2,1/2), (3/2,1/2,-1/2), and (3/2,1/2,-1/2,-3/2), respectively. Region (iii) is then similar to the case of a pure Zeeman splitting.  Hence, at the crossovers the character of the bound states changes. The corresponding chemical potentials $\mu_i$ in Eqs. (\ref{var0})-(\ref{var3}) are then 
\vskip 0.1in
\par\noindent
Region (i) ($H \le 3D$)
\par\noindent
\begin{tabular}{lll}
$\mu_0=\mu+3D+3H/2$, & & $\mu_1=\mu+3D$ \\
$\mu_2=\mu+D+H/6$, & & $\mu_3=\mu$ 
\end{tabular}
\vskip 0.1in
\par\noindent
Region (ii) ($3D \le H \le 6D$)
\par\noindent
\begin{tabular}{lll}
$\mu_0=\mu+3D+3H/2$, & & $\mu_1=\mu+H$ \\
$\mu_2=\mu+D+H/6$, & & $\mu_3=\mu$ 
\end{tabular}
\vskip 0.1in
\par\noindent
Region (iii) ($6D \le H$)
\par\noindent
\begin{tabular}{lll}
$\mu_0=\mu+3D+3H/2$, & & $\mu_1=\mu+H$ \\
$\mu_2=\mu-D+H/2$, & & $\mu_3=\mu$ .
\end{tabular}
\vskip 0.1in
\par\noindent
Similarly one can obtain the chemical potentials for $D < 0$. The procedure is completely analogous to that used for magnetic impurities (degenerate Anderson model in the $U \to \infty$ limit with Zeeman and crystalline field splittings).\cite{Schl89}

The phase diagram for $D=1/3$ and $D=-1/3$ is shown in Figs. 4 and 5, respectively. For these parameters the level crossings are at $H=1$ and $H=2$. The level crossings stabilize the I+III+IV and I+II+IV (II+III+IV) mixed phases over the I+II+III+IV mixed phase.  As discussed above, the phase I+III+IV for $H \geq 1$ involves different condensates than for $H \leq 1$. All phase boundaries are the consequence of one of the four rapidity bands getting empty and, hence, a transition involves a one-dimensional van Hove singularity with the corresponding consequences on the density of states and low-$T$ specific heat. For small magnetic fields the phase is a mixture of two-particle and four-particle bound states. For larger fields $H$, the four-particle bound states are only favorable if the density of particles is high enough. Also for intermediate magnetic fields the phase diagram for $D>0$ is very different from that of $D<0$.

\section{Conclusions}

We studied an ultracold gas of fermionic atoms with an attractive contact potential by solving the corresponding Bethe {\it ansatz} equations. We obtained the phase diagram  for a $S=3/2$ in a magnetic field ($\mu$ vs. $H$) within the grand canonical ensemble. Four elementary states can occur: (i) polarized unbound atoms with spin-component $S_z=3/2$, (ii) bound pairs of atoms with spin-components $S_z=3/2$ and $S_z=1/2$, (iii) bound states of three particles with spin-components $S_z=3/2$, $S_z=1/2$ and $S_z=-1/2$, and (iv) bound states of four particles, one with each spin-component. Mixed phases of different classes of bound states dominate the phase diagram. For a given chemical potential the phases are homogeneous and display no long-range order. The transitions between phases are crossovers of the Prokovskii-Talapov type.

There are several advantages of working in the grand-canonical ensemble vs. the canonical ensemble,\cite{Guan2} where the number of particles is kept fixed. (1) By rescaling all quantities in the integral equations for the dressed energy potentials, $\varepsilon^{(q)}$, with the interaction strength $|c|$, one obtains {\it universal} equations for the phase diagram $\mu$ vs. $H$. Our phase diagram shown in Figs. 1 and 2 is then valid for all attractive $|c|$. This is not the case if the total number of particles is kept fixed. (2) Since the diameter of the tube gradually changes with position from the center of the trap to its boundaries, the effective local chemical potential varies along the tube. Within the local density approximation this change can be represented by a harmonic potential and as a consequence of the $x$-dependence of $\mu$ there is an inherent tendency of phase separation,\cite{Liao} i.e. the trap is inhomogeneous. At different positions of the trap then different phases may be realized and a sequence of transitions should be observed along the trap. (3) Josephson tunneling between tubes and interactions between particles in different tubes,\cite{Yang,Parish} may give rise to a dimensional crossover from one-dimension to a higher dimension. This gives rise to long-range order of quantities that are generalizations of Cooper pairs for $S=1/2$ to higher spin. The system remains strongly anisotropic and pure (there are almost no impurities) and is hence favorable for inhomogeneities like modulations of the order parameter of the FFLO type in the presence of an external magnetic field. All these properties can only be studied within the framework of the grand-canonical ensemble.    

We also investigated the theoretically interesting situation of a quadrupolar splitting superimposed with the Zeeman field. In this case, the spin energy levels display two crossovers as a function of the magnetic field. Hence, the character of the bound states before and after the crossover changes. In general bound states of four particles are not favorable for low particle density or high magnetic fields.

The phase diagram for larger spin values can be obtained the same way but the calculations and the results are considerably more involved and complicated. An extension to hard-core bosons (particles with integer spin) is also possible,\cite{Guan1,Cao,Chen} since the Bethe {\it ansatz} equations for hard-core bosons are the same ones as for fermions. Note that wave functions for hard-core bosons have a node when two particles are at the same position, in analogy to what is required by Pauli's principle for fermions. 

\section*{Acknowledgments}
P.S. would like to thank Dr. Ed Myers for a helpful discussion about the experimental situation. P.S. is supported by the U.S. Department of Energy under grant DE-FG02-98ER45707. 


\begin{thebibliography}{99}
\bibitem{He} This is known for nucleons that have a pseudospin in addition to the spin; $\alpha$-particles are bound states of four nucleons, the nuclei of $^3$He and tritium are examples of bound states of three nucleon and the nucleus of deuterium binds two nucleons.
\bibitem{vanHove} Within the grand canonical ensemble the dressed energy potentials are all quadratic in the rapidity as seen in Eqs. (\ref{var0})-(\ref{var3}). The solution of these equations yields a one-dimensional van Hove singularity with a square root divergence when a band is emptied (Prokovskii-Talapov transition). This is the situation relevant to this paper. We believe that this remains true within the canonical ensemble (with constrained number of particles),\cite{Schl91} but this point is contested in Refs. [\onlinecite{BW}].
\end{thebibliography}
\end{document}